\documentstyle[12pt]{article}
\begin{document}

\newcommand{\D} {\Delta}
\newcommand{\Db} {\bar \Delta}
\newcommand{\be}{\begin{equation}}
\newcommand{\ee}{\end{equation}}
\newcommand{\ba}{\begin{eqnarray}}
\newcommand{\ea}{\end{eqnarray}}
\def\pct#1{\input epsf \centerline{ \epsfbox{#1.eps}}}
\begin{titlepage}
\hbox{\hskip 13cm ROM2F-95/18  \hfil}
\hbox{\hskip 13cm \today \hfil}
\vskip 4cm
\begin{center}

{\Large  \bf    Some \ Properties \ of \ Open - String \ Theories}

\vspace{2.5cm}

{\large Augusto \ Sagnotti}

\vspace{0.8cm}

{\sl Dipartimento di Fisica\\
Universit\`a di Roma \ ``Tor Vergata'' \\
I.N.F.N.\ - \ Sezione di Roma \ ``Tor Vergata'' \\
Via della Ricerca Scientifica, 1 \ \
00133 \ Roma \ \ ITALY}
\vspace{3.5cm}
\end{center}

\abstract{
Open-string theories may be related to suitable models of oriented closed
strings. The resulting construction of ``open descendants'' is illustrated in
a few simple cases that exhibit some of its key features.}
\vskip 4 cm
\begin{center}
{\large \sl Presented \ at \ SUSY \ '95, Palaiseau, FRANCE, May 1995}
\end{center}
\vfill
\end{titlepage}
$ \ $
\vskip 3.5cm
\begin{center}
{\large  \bf  Some \ Properties \ of \
Open - String \ Theories}

\vspace{6.0ex}

{\large Augusto \ Sagnotti}

\vspace{1.5ex}

{\large \it Dipartimento di Fisica\\
Universit{\`a} di Roma \ ``Tor Vergata'' \\
I.N.F.N.\ - \ Sezione di Roma \ ``Tor Vergata'' \\
Via della Ricerca Scientifica, 1 \ \
00133 \ Roma \ \ ITALY}

\vspace{4.5ex}

{\bf Abstract}
\end{center}

Open-string theories may be related to suitable models of oriented closed
strings. The resulting construction of ``open descendants'' is illustrated in
a few simple cases that exhibit some of its key features.

\vspace{3.0ex}
\begin{flushleft}
{\large \bf Introduction}
\end{flushleft}

The key features of (perturbative) spectra and interactions for models
of oriented closed strings have been clarified by a number of groups
\cite{closed}
at the end of the last decade.
These works, all  inspired by ref. \cite{gso},
have the virtue of providing general, handy rules for constructing
closed-string
spectra from a variety of conformal field theories.  Basically, any
modular invariant combination of conformal field theories that respects the
spin-statistics relation between bosonic and fermionic contributions
to the vacuum
amplitude and saturates the total conformal anomaly defines the perturbative
spectrum of a model of oriented closed strings.  The resulting plethora of
solutions, a nice arena for string models of particle physics,
is actually rather disturbing from the viewpoint of string unification.
It is fortunate in this respect that all
these constructions are tied to string perturbation theory, for which no
genuine weak-coupling arguments are available. Recent work on string
dualities, inspired to a large extent by general features of the low-energy
supergravity and reviewed in a number of contributions
to this volume, is bringing new, concrete
evidence for the long-held feeling that all string theories are somehow
different manifestations of a single underlying entity.

The purpose of this talk is to illustrate the state of the art of a program
aimed at associating ``open descendants'' to suitable models of oriented
closed strings and, more generally, to suitable conformal field theories.  This
started in the late eighties with the proposal of ref. \cite{car} and is
surely lagging behind the corresponding work for models of oriented closed
strings, but has by now
evolved into an algorithm capable of associating to suitable closed spectra
additional open spectra with well-defined patterns for the breaking of the
internal (Chan-Paton) symmetry \cite{cp}.  I shall try to describe
this setting by referring to a few simple cases capable of displaying some of
its key
features.  Technical details may be found in the original papers \cite{bs}
\cite{bps1} \cite{bps2} \cite{as} and in
a forthcoming review article.  These open-string
models supplement the existing ``oriented closed'' zoo, while providing
string-based
descriptions for some additional classes of  (super)gravity models.  The recent
work on
string dualities reviewed in other contributions to these Proceedings suggests
possible
applications to the strong-coupling regime of other string models.

For simplicity, I shall confine my attention to a few key
properties of genus-one partition functions that have been
investigated in some detail.  The
simplest open-string models are the descendants of the Type-IIb superstring
and of the two ten-dimensional tachyonic
models first introduced in ref. \cite{sw}. In writing their vacuum amplitudes,
I shall omit the modular integrations, while trading the theta
constants for the four characters of level-one $SO(8)$ representations.
These are simply related to one another, since
\ba
& &O_8 = 	{1 \over {2 \eta^4}} \biggl( \ \theta^4 \biggl[ {0 \atop 0} \biggr]
\ + \
\theta^4 \biggl[ {0 \atop 1/2} \biggr] \ \biggr)
\quad \ , \qquad V_8 =	{1 \over {2 \eta^4}} \biggl( \ \theta^4
\biggl[ {0 \atop 0} \biggr] \ - \
\theta^4 \biggl[ {0 \atop 1/2} \biggr] \ \biggr) \quad , \nonumber \\
& &S_8 = 	{1 \over {2 \eta^4}} \biggl( \ \theta^4 \biggl[ {1/2 \atop 0} \biggr]
 \ + \
\theta^4 \biggl[ {1/2 \atop 1/2} \biggr] \ \biggr) \ ,
\qquad C_8 =	{1 \over {2 \eta^4}} \biggl( \ \theta^4 \biggl[ {1/2 \atop 0}
\biggr] \ - \
\theta^4 \biggl[ {1/2 \atop 1/2} \biggr] \ \biggr) 	\ ,
\label{so8}
\ea
with $\eta$ the Dedekind function,
but the $SO(8)$ characters have the additional virtue of providing
an orthogonal decomposition of the spectrum.
In this notation, the interesting ten-dimensional models correspond
to the partition functions
\ba
& &T_{IIb} \ = \ {|V_8 - S_8 |}^2 \quad , \nonumber \\
& &T_{0a} \ = \ |O_8 |^2 + |V_8 |^2 + S_8 {\bar{C}}_8 + C_8 {\bar{S}}_8 \quad ,
\nonumber \\
& &T_{0b} \ = \ |O_8 |^2 + |V_8 |^2 + |S_8 |^2 + |C_8 |^2 \quad .
\label{tendim}
\ea
All other ten-dimensional models have a left-right asymmetric spectrum,
including
the type-IIa superstring, whose partition function in this notation
reads
\be
T_{IIa} \ = \ ( V_8 - S_8 ) ( {\bar{V}}_8 - {\bar{C}}_8 )	\quad .
\label{type2a}
\ee
\vskip 24pt
\begin{flushleft}
{\large \bf The Klein-bottle Projection and the Crosscap Constraint}
\end{flushleft}

The starting point in the construction of open descendants is a projection
of the closed spectrum mixing left and right movers.  This is
attained supplementing the modular invariant (halved) torus partition function
with a Klein-bottle contribution.
\vskip 7pt
\input epsf \centerline{ \epsfbox{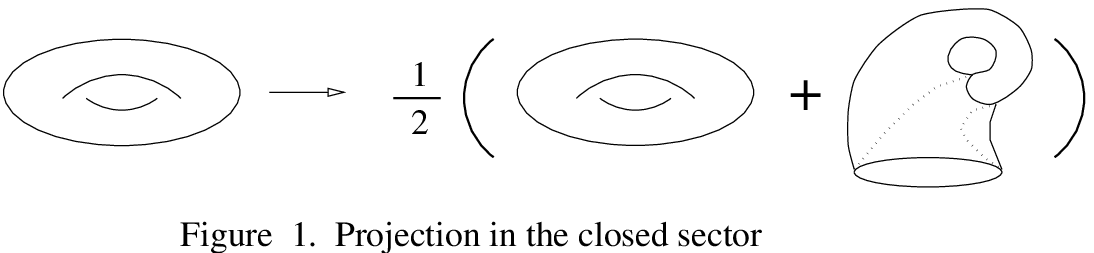}}
\vskip 7pt
\noindent
In the simplest of possible settings, the
bosonic string, this operation is what in the early days of String Theory
was said to lead from the ``extended'' to the
``restricted'' Shapiro-Virasoro model.  In the ``extended'' model, all states
in, say, the light-cone description are obtained acting on the vacuum
with polynomials in the transverse left-moving oscillators ${\alpha_n}^i$
and in the transverse right-moving oscillators ${ {\tilde{\alpha}}_n}^i$,
with the familiar Virasoro constraint on the
entity of the corresponding total excitations. On the other hand, in the
``restricted'' model only polynomials with an additional discrete symmetry
under the interchange of all the ${\alpha_n}$ with the ${\tilde{\alpha}}_n$ are
allowed.
In Superstring Theory, anomalies and
their manifestation in this context, tadpoles \cite{pc},
grant the motivation for the Klein-bottle projection that,
in a general conformal field theory, exposes the rich structure of open
models, exhibiting their patterns of internal symmetry and their planar duality
in
full-fledged form.

Let me follow the steps leading to the construction of open
descendants, while referring to the models described in the previous Section.
First of all, one is to halve the
torus contribution, but I shall refrain from doing so explicitly.  Then
one is to add the Klein-bottle contribution.  This is the first crucial step,
since
it determines the symmetry of the various sectors of the
``restricted'' closed spectrum under the interchange of
their ``left'' and ``right'' parts.
In general, the choice of Klein-bottle  projection is {\it not} unique
\cite{wzw1}
\cite{wzw2} while, technically, the various projections differ in the
signs for the characters that appear diagonally in the torus
amplitude, that determine the (anti)symmetry of the restricted spectrum. For
instance, the
type-IIb model of eq. (\ref{tendim}) could in principle admit four Klein-bottle
projections, corresponding to all available choices of signs for the two
characters $V_8$ and $S_8$. Actually, the fusion algebra introduces
strong restrictions.  Since in general only
(anti)symmetric combinations of {\it all} left and right Verma modules are
compatible with a modular invariant torus amplitude,
it is simple to understand how these restrictions
arise by ``fusing'' pairs of states.  Thus, in this case the fusion of
two (space-time) spinors yields a vector, and the $NS \otimes NS$
sector must always be symmetrized. The
lesson that may be drawn from a closer inspection of this simple example is
actually
quite general: the available choices correspond to the $Z_2$ automorphisms of
the fusion
algebra compatible with the torus $GSO$ projection.  With this proviso, one can
see that
the ten-dimensional models of eq. (\ref{tendim}) allow at most {\it four} types
of
Klein-bottle projection, though two of them are really equivalent.  The first,
most natural
choice, corresponds to the natural basis of characters for the space-time
Lorentz group
$SO(1,9)$, namely
$( O_8, V_8, - S_8, - C_8 )$, and results in the projections
\ba
& &K_{IIb} = {1 \over 2} \ \bigl( V_8 - S_8 \bigr) \quad , \nonumber \\
& &K_{0a} = {1 \over 2} \ \bigl( O_8  + V_8 \bigr) \quad , \nonumber \\
& &K_{0b} = {1 \over 2} \ \bigl( O_8  + V_8 - S_8  - C_8 \bigr) \quad ,
\label{ktendim}
\ea
whereby $NS-NS$ sectors are symmetrized while $R-R$ sectors are
antisymmetrized.  The other choices correspond to the basis $( O_8, V_8, S_8,
C_8
)$, to the basis
$( - O_8, V_8, -S_8, C_8 )$ and to the basis $( - O_8, V_8, S_8, -C_8 )$.
Only the first Klein-bottle projection is allowed in the type-IIb superstring,
while
the last two, clearly equivalent after a parity transformation,
change the relative weight of the $RR$ sectors, and are therefore
incompatible with the $GSO$ projection of the $Oa$ model as well.  On the other
hand, in the
$Ob$ model one is lead to the additional genuinely inequivalent Klein-bottle
projections
\ba
& &K^{\prime }_{0b} = {1 \over 2} \ \bigl( O_8  + V_8 + S_8  + C_8 \bigr) \quad
, \nonumber
\\ & &K^{\prime \prime}_{0b} = {1 \over 2} \ \bigl( - O_8  + V_8 + S_8  - C_8
\bigr)
\quad .
\label{kntendim}
\ea

It is instructive to take a closer look at the lowest-mass states of the $Ob$
descendants resulting from the three types of projection.
Standard properties of the ten-dimensional Lorentz group \cite{gsw} then imply
that
$K_{0b}$ leaves a tachyon, a graviton and a dilaton in the $NS \otimes NS$
sectors, as well as two more scalars and two more antisymmetric tensors in
the $R \otimes R$ sectors, and thus projects out all ``chiral'' closed-string
fields.  On the other hand, $K^{\prime}_{0b}$ also leaves a tachyon, a graviton
and a dilaton in the $NS \otimes NS$ sectors, but now the the $R \otimes R$
sectors
contain both a self-dual and an antiself-dual four-form, with a resulting
projected
closed spectrum that is again not chiral.
The last projection, $K^{\prime \prime }_{0b}$, is actually the most
interesting one.
Indeed, the corresponding  closed spectrum does not contain a tachyon, but only
a
graviton and an antisymmetric tensor in the $NS \otimes NS$ sectors, and
is chiral, since in the $R \otimes R$ sectors it contains a self-dual
four-form,
as well as an antisymmetric tensor and a scalar.  As
we shall see, there are chiral open descendants of the $Ob$ model
where the Green-Schwarz
\cite{gs} mechanism eliminates all gauge and gravitational anomalies.  On the
other hand, according to our rule, both the
$Oa$ and the $IIb$ models allow only one Klein-bottle projection.  The former
leads to
the models of ref. \cite{bs}, while the latter leads to the Type-I $SO(32)$
superstring.
Many of these remarks apply to arbitrary conformal models symmetric under the
interchange of their left and right parts.
Ref. \cite{bps1} contains a discussion of the open descendants of $AA$ minimal
models with a totally symmetric Klein-bottle projection, while refs.
\cite{wzw1}
and \cite{wzw2} contain several details on the general case for $SU(2)$ $WZW$
models.
\vskip 7pt
\input epsf \centerline{ \epsfbox{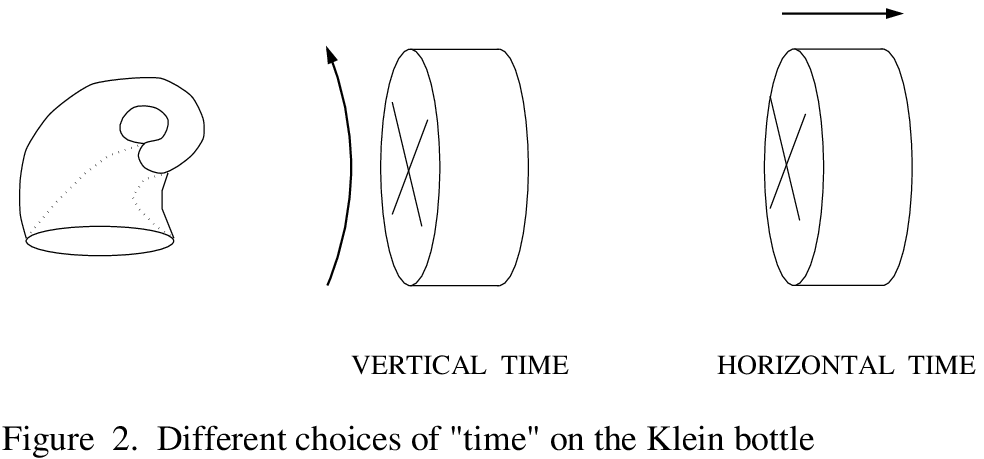}}
\vskip 7pt

There is actually a nice geometrical setting for the various phase choices
in the Klein-bottle projections.  Indeed, a Klein
bottle may be pictured as a self-intersecting surface or,
alternatively, as a tube terminating at two crosscaps.  A crosscap, or real
projective plane, is a simple instance of non-orientable surface free of
boundaries,
and may be regarded as a disk where all pairs of opposite  points lying along
the boundary
are identified.
\vskip 7pt
\input epsf \centerline{ \epsfbox{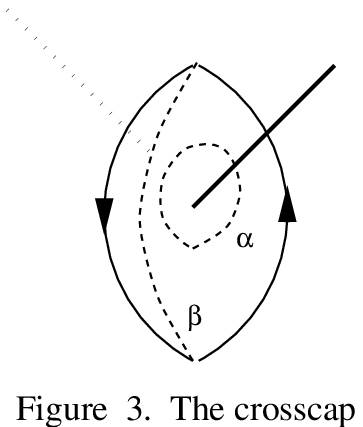}}
\vskip 7pt
\noindent
Given a two-point amplitude as in fig. 3, moving one
of the two punctures
probes the fundamental group of the simply-punctured surface.  According to the
familiar
properties of closed-string models, no price is to be paid when transporting
one puncture
around the other (path $\alpha$). However, here there is a more elementary
move, namely the
puncture may be displaced up to a point to emerge from the other identified
point (path
$\beta$).  Since this move squares to the previous one, one is left with some
free signs
that reflect the behavior of the various two-point functions upon transport
around the
crosscap. The free signs are  actually to be
compatible with the fusion rules, and correspond to the free signs in the
Klein-bottle
projection \cite{wzw1} \cite{wzw2}.

Before turning to the open spectrum, let me discuss in some detail the other
available
choice of ``time'' for the Klein bottle.  Referring to fig. 2, the previous
choice
may be termed ``vertical'', while this new choice may be termed ``horizontal''.
 The
two are related by the familiar modular transformation $\tau_2 \rightarrow
1/{\tau_2}$, and the second type of amplitude, that I shall denote $\tilde{K}$,
exhibits the propagation of the {\it closed} spectrum between a pair of
crosscaps.
One should therefore demand that, in terms of the proper basis of characters,
{\it
all} coefficients in the transverse Klein-bottle amplitude be positive.  This
is
actually guaranteed rather nicely by our previous rule. Thus, for instance, for
the
descendants of the ten-dimensional $Ob$ model,
\ba
& &\tilde{K}_{0b} = {2^6 \over 2} \ V_8 \qquad , \nonumber \\
& &\tilde{K}^{\prime }_{0b} = {2^6 \over 2} \ O_8 \qquad , \nonumber \\
& &\tilde{K}^{\prime \prime}_{0b} = \ - \ {2^6 \over 2} \ C_8  \qquad ,
\label{ktildetendim}
\ea
and, as anticipated, all coefficients are indeed positive if referred to the
natural basis of
space-time characters, $( O, V, -S, -C )$.  It
should be appreciated that these coefficients have a nice physical
interpretation: they are squared moduli of the normalizations for the one-point
functions of
primary fields in front of a crosscap.  Their values are determined to a large
extent by the crosscap constraint \cite{cap} \cite{wzw1}
\cite{wzw2}, an eigenvalue equation expressing the
equivalence of the different sewings for the amplitude of fig. 3, and fully
so for all $SU(2)$ $WZW$ models in the $ADE$ series.

In our ten-dimensional models, the Klein-bottle projections $K$ and the
corresponding
vacuum-channel amplitudes $\tilde{K}$ are related by the $S$ matrix
\be
S \ = \ {1 \over 2}
\pmatrix{1&1&1&1& \cr 1&1&-1&-1 \cr 1&-1&1&-1 \cr 1&-1&-1&1}
\qquad ,
\label{stendim}
\ee
determined by the familiar properties of theta functions.
The overall powers of two in the vacuum-channel amplitudes of eqs.
(\ref{ktildetendim})
deserve some additional comments since, as we shall see, they are related to
the size of
the open-string Chan-Paton groups. They draw their origin from a natural
prescription to
deal with divergent infrared contributions to the vacuum channel, whereby {\it
all} vacuum
amplitudes are expressed in terms of the moduli of their double covers.  This
prescription
\cite{gs2} \cite{cp} allows a convenient comparison between the contributions
of the three
surfaces of zero Euler character that determine the spectrum of open-string
models, namely
the Klein bottle, the annulus and the M\"obius strip, and is equivalent
\cite{bs1} to other methods of dealing with tadpoles and divergences
\cite{dgw}.
\vskip 24pt
\begin{flushleft}
{\large \bf The Open Sector and Chan-Paton Symmetry Breaking}
\end{flushleft}

The next step in the construction has to do with the new, ``twisted'', sector.
This
involves open strings, and a new projection obtained adding to a (halved)
annulus amplitude a M\"obius contribution.
\vskip 7pt
\input epsf \centerline{ \epsfbox{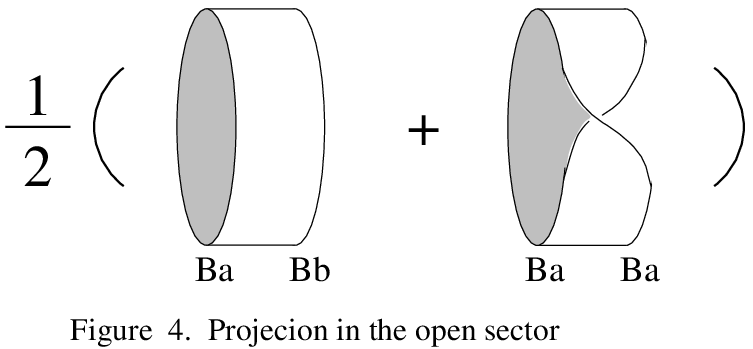}}
\vskip 7pt
\noindent
These last two surfaces have vanishing
Euler character and thus contribute to the same order of perturbation theory
but,
differently from the Klein bottle, have respectively two boundaries and one
boundary
and one crosscap.  The
boundaries may be pictured as drawn by the ends of open strings, the site of
their Chan-Paton charges \cite{cp}.  Therefore, it should come as no surprise
that these
additional contributions are polynomials of degree two and one in the
multiplicities of the various charge spaces.  The structure of these
polynomials is
subject to strong constraints, since it is to be
compatible with the factorization of disk amplitudes.
\vskip 7pt
\input epsf \centerline{ \epsfbox{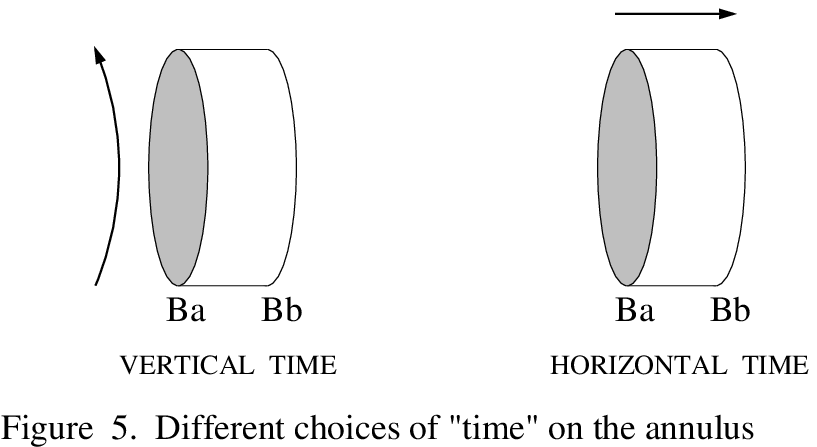}}
\vskip 7pt
\noindent
Moreover, since a ``horizontal'' time in the annulus
amplitude exhibits a vacuum channel where the closed spectrum propagates
between a pair
of holes (fig. 5), the coefficients in $\tilde{A}$ must satisfy positivity
constraints with respect to a proper basis of characters, in a similar fashion
to what we
have seen for $\tilde{K}$.  These coefficients are then properly
interpreted as squared moduli of the normalizations of one-point functions on
the disk.

In building the annulus amplitude, it is very useful to begin from the vacuum
channel. This
corresponds to the ``horizontal time'' in fig. 5 and accommodates the
propagation of the
{\it closed} spectrum, that is by now under control for a given model.  One is
thus faced
with the choice of a number of (squared moduli of) reflection  coefficients,
and in some
cases, including the simple ones I am dealing with here, the proper choice is
fully
determined by symmetry. The basic rule is simple to state.  Given a closed
spectrum
described by a (quasi)diagonal partition function of the type
\be
T \ = \ \sum_{i,j} \ N_{ij} \ \chi_i \ {\bar{\chi}}_j	\qquad ,
\label{torus}
\ee
the characters $\chi_i$ that may flow in vacuum channel of the annulus are
those {\it paired
with their conjugates with respect to the given symmetry by the closed-string
$GSO$
projection} \cite{bs}. This corresponds to the intuitive idea that a state
flowing, say, to
the right is to be turned, upon reflection, into its conjugate (in the sense of
the
closed-string
$GSO$ projection). If the boundary is to
respect a given (extended) symmetry, a non-trivial reflection is possible only
if the
$GSO$-conjugate happens to be also conjugate with respect to the given
symmetry.  All
this is perhaps simpler to appreciate by referring to our examples.  Thus, in
the $IIb$ case
there is a single (super)sector, corresponding to $V_8 - S_8$, and
\be
{\tilde{A}}_{IIb} \ = \ {2^{-5} \over 2} \ {\alpha^2}_{IIb} \ \bigl( V_8 - S_8
\bigr)
\qquad ,
\label{ant2b}
\ee
where the overall normalization is chosen for later convenience.
On the other hand, in the $O_a$ model only $O_8$ and $V_8$ are paired with
their conjugates
by the projection of eq. (\ref{tendim}), since both $S_8$ and $C_8$ are
self-conjugate in
$SO(8)$, and therefore
\be
{\tilde{A}}_{0a} \ = \ {2^{-5} \over 2} \bigl( \ {\alpha^2}_{0a} \ V_8 \ + \
{\beta^2}_{0a}
\ O_8 \ \bigr)
\qquad .
\label{ant0a}
\ee
Finally, in the $Ob$ model all characters are paired with
their conjugates by the closed-string $GSO$ projection, and the annulus vacuum
channel
contains four reflection coefficients:
\be
{\tilde{A}}_{0b} \ = \ {2^{-6} \over 2} \bigl( \ {\alpha^2}_{0b} \ V_8 \ + \
{\beta^2}_{0b}
\ O_8  \ - \  {\gamma^2}_{ob} \ S_8 \ - \ {\delta^2}_{0b} \ C_8 \ \bigr) \qquad
{}.
\label{ant0b}
\ee

The number of independent\footnote{In more complicated models the
coefficients may turn out not to be independent.  More details on this issue
may be found
in ref. \cite{wzw2}.} coefficients is a very
important datum in the construction since, as we shall see, it determines the
number of
charge sectors in the models.  Moreover, since the M\"obius strip may be seen
as a tube with
a hole and a crosscap at the ends, the vacuum M\"obius channel
$\tilde{M}$ is also determined by the two amplitudes $\tilde{K}$ and
$\tilde{A}$.  Thus,
in our models,
\ba
& &{\tilde{M}}_{IIb} \ = \ {2 \over 2} \ \alpha_{IIb} \
\bigl( V_8 - S_8 \bigr) \qquad , \nonumber \\
& &{\tilde{M}}_{0a} \ = \ {2 \over 2} \biggl( \ \alpha_{0a} \ V_8 \ + \
\beta_{0a}
\ O_8 \ \biggr) \qquad , \label{moebius} \\
& &{\tilde{M}}_{0b} \ = \ {2 \over 2} \left( \ \alpha_{0b} \ V_8 \ + \
\beta_{0b} \ O_8  \ - \  \gamma_{ob} \ S_8 \ - \ \delta_{0b} \ C_8 \ \right)
\qquad .
\nonumber
\ea
In these expressions the overall coefficients are geometric means of those in
$\tilde{K}$
and $\tilde{A}$, with an additional factor of two that
reflects the combinatorics of the M\"obius vacuum channel.  Once more, this
choice may be
justified by a careful analysis of the measure, and draws its motivation from
the
prescription of referring all amplitudes to their double covers.

One may now turn the $\tilde{A}$ amplitudes into direct-channel open-string
partition
functions using the $S$ matrix of the previous section.  On the other hand, a
similar
operation for the M\"obius amplitudes requires the use of a different matrix,
$P$,
related to the two more familiar matrices $S$ and $T$ by
\be
P \ = \ T^{1/2} \ S \ T^2 \ S \ T^{1/2} \qquad .
\label{pmat}
\ee
Rather than performing these transformations directly, I would like to turn
momentarily to
the direct channel, in order to exhibit an ansatz for the Chan-Paton charge
assignments.  This  choice drew
its motivation from previous work of Cardy \cite{cardy}. Aiming
at a more direct derivation of the Verlinde formula \cite{verlinde}, he had
considered the
annulus amplitude with fixed boundary conditions in a generic diagonal rational
conformal
field theory, providing a novel interpretation of the fusion-rule coefficients
$N_{ij}^k$: they determine the bulk content ($k$) corresponding to boundary
conditions
$i$ and
$j$.  The observation of ref. \cite{bs} is that, given the one-to-one
correspondence between
bulk and boundary (charge) sectors of these rational models, the ansatz may be
turned into
one for an annulus amplitude with broken Chan-Paton symmetry, namely
\be
A \ = \ {1 \over 2} \ \sum_{i,j,k} \ N_{ij}^k \ n^i \ n^j \ \chi_k \qquad .
\label{ansatz}
\ee
Actually, the Verlinde formula makes this expression particularly appealing,
since it
implies that
\be
\tilde{A} \ \sim \ {1 \over 2} \ \sum_{i} \ \chi_i \left( \sum_{j,k}
{{S_{jk} \ n^k} \over {\sqrt{S_{j0}}}} \right)^2 \qquad ,
\ee
and therefore {\it all} coefficients in $\tilde{A}$ are perfect squares, as
demanded by
the structure of this contribution.  We have developed
a foolproof procedure to construct the charge assignments in general models
\cite{wzw1}
\cite{wzw2}, and indeed in the non-diagonal $WZW$ models eq. (\ref{ansatz})
does not apply,
though the simple models considered in this talk may be constructed in way
directly
suggested by this observation. Before displaying the corresponding amplitudes,
let me
introduce another, related, concept.  This has to do with ``complex''
Chan-Paton charges,
and is a generalization of a long-known property of boundaries, that in
oriented open
strings (or, in our case, in oriented sectors of open spectra) carry arrows
that account for
their orientation.  Technically, ``complex'' boundaries carry charges valued in
fundamental
representations of unitary groups, whereas ``real'' boundaries carry charges
valued in
fundamental representations of symplectic or orthogonal groups \cite{cp}.  In
these models,
this peculiarity presents itself whenever a naive transverse-channel amplitude
would
violate positivity, as will be seen in a simple example below. The further
restrictions that
restore consistency are, basically, the numerical constraints $n = \bar{n}$ for
pairs of
conjugate charges. ``Complex'' charges, however, appear also when some
characters are not
self conjugate \cite{bs}.

Turning to the explicit form of the amplitudes, let me proceed by considering
the
descendants of the type-IIb model.  These are type-I models, and it has long
been known
that only one of them is anomaly free \cite{gs}.  In order to see how the
restrictions
manifest themselves, let me write explicitly all the amplitudes:
\ba
& &A_{IIb} \ = \  {n^2 \over 2} \bigl( \ V_8 \ - \ S_8 \ \bigr) \qquad ,
\nonumber \\
& &M_{IIb} \ = \ - \ {n \over 2} \bigl( \ V_8 \ - \ S_8 \ \bigr) \qquad ,
\nonumber \\
& &{\tilde{A}}_{IIb} \ = \ 2^{-5} \ {n^2 \over 2} \bigl( \ V_8 \ - \ S_8 \
\bigr)
\qquad , \nonumber \\
& &{\tilde{M}}_{IIb} \ = \ - 2 \ {n \over 2} \bigl( \ V_8 \ - \ S_8 \ \bigr)
\qquad .
\label{openIIb}
\ea
The overall sign of the M\"obius terms is conventional at this stage.  A
positive $n$ would
imply an orthogonal gauge group, with  $n(n-1)/2$ gauge bosons, while a
negative $n$
would imply a symplectic gauge group, with  $|n|(|n|+1)/2$ gauge bosons, and
the proper choice
is determined by a tadpole condition.  The precise relation between tadpole
conditions and
anomaly cancellations was first pointed out for the type-I superstring in ref.
\cite{pc}:
the anomalies are linked to tadpoles of {\it massless unphysical states}.  In
our setting it
is simple to track such states, and there is one in these models: it
corresponds to the
sector
$S_8$ in the vacuum channel, that would start with a massless $RR$ scalar
projected out of the closed spectrum by the Klein-bottle of
eq. (\ref{ktendim}).  Setting to zero its total one-point function at genus
one-half,
determined by eqs. (\ref{ktildetendim}) and (\ref{openIIb}),
yields a quadratic equation for $n$, that by construction is guaranteed to have
two
coincident roots, namely (fig. 6)
\be
2^{10} \ + \  n^2 \ - \ 2^6 n \ = \ 0 \qquad ,
\label{tadIIb}
\ee
where the three terms originate from $\tilde{K}$, $\tilde{A}$ and $\tilde{M}$.
As
anticipated, the unique solution, $n=32$, leads to a unique anomaly-free
$SO(32)$ model, a
long-known result of Green and Schwarz \cite{gs}.
\vskip 7pt
\input epsf \centerline{ \epsfbox{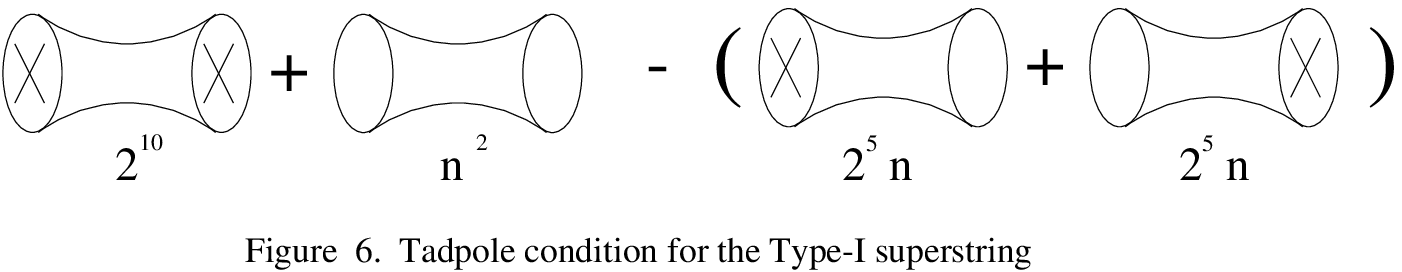}}
\vskip 7pt

Let me now turn to the $Oa$ models, that have a unique Klein-bottle projection
and are thus
simpler to deal with.  In this case the direct-channel annulus has two charge
sectors, and
one has the consistent set of assignments
\ba
& &A_{0a} \ = \  {{{n_B}^2 + {n_F}^2} \over 2} \bigl( \ O_8 \ + \ V_8 \ \bigr)
- n_B n_F \ \bigl( \ S_8 \ + \ C_8 \ \bigr) \quad , \nonumber \\
& &M_{0a} \ = \  {{n_B + n_F} \over 2} \bigl( \ O_8 \ - \ V_8 \ \bigr)   \quad
, \nonumber \\
& &{\tilde{A}}_{0a} \ = \ {{2^{-5}} \over 2} \ \left( \ {\bigl( n_B + n_F
\bigr)}^2 \ V_8 \ +
\ {\bigl( n_B - n_F \bigr)}^2 \ O_8 \right) \quad , \nonumber
\\
& &{\tilde{M}}_{0a} \ = \ - 2 \ {{n_B + n_F} \over 2} \bigl( \ O_8 \ + \ V_8 \
\bigr) \quad ,
\label{open0a}
\ea
that determine the coefficients in eq. (\ref{ant0a}).  This simple class of
models is
interesting, since it exhibits a
new phenomenon, Chan-Paton symmetry breaking.
\vskip 7pt
\input epsf \centerline{ \epsfbox{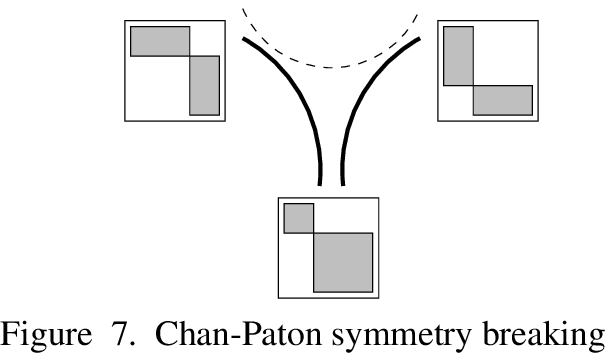}}
\vskip 7pt
\noindent
Referring to fig. 7 one can appreciate
that allowing, say, $n_F \neq 0$, has the effect of emptying the Chan-Paton
charge matrices
of the gauge vectors while moving their charges to new sectors of the spectrum.
Moreover, this
is all manifestly compatible with factorization.
Thus, for instance, if $n_B$ and $n_F$ are both positive there are gauge bosons
(corresponding to $V_8$) in the adjoint representation of  $SO(n_B ) \otimes
SO(n_F )$,
tachyons (corresponding to $O_8$) in the symmetric two-tensor and scalar
representations of
the same group, and Majorana fermions in the $( n_B , n_F )$ representation.
This spectrum
is manifestly non-chiral, both in the closed sector projected according to eq.
(\ref{ktendim}) and in the open sector. Correspondingly, the only massless
tadpole in the
vacuum channel corresponds to the
$V_8$ sector, a physical one, and determines the dilaton tadpole at genus
$1/2$.  Thus here
the tadpole condition, that would fix the total size of the (orthogonal) gauge
group to $n_B
+ n_F = 32$, is {\it not} related to gauge and gravitational anomalies as in
the type-IIb
case. To date, the role of these additional tadpole conditions is
not understood to my own satisfaction.

Let me now turn to the $Ob$ models, beginning with the usual ansatz, that in
this case reads
\ba
A_{0b} \ &=& \  {{{n_o}^2 + {n_v}^2 + {n_s}^2 + {n_c}^2} \over 2}  \ V_8 \ +
\bigl( n_o n_v + n_s n_c \bigr) \ O_8 \nonumber \\ &-& \ \bigl( n_o n_c + n_v
n_s \bigr) \ S_8
\ - \ \bigl( n_o n_s + n_v n_c \bigr) \ C_8 \quad , \nonumber \\
M_{0b} \ &=& \  - \ {{n_o + n_v + n_s + n_c } \over 2}  \ V_8   \quad ,
\nonumber
\\
{\tilde{A}}_{0b} \ &=& \ {{2^{-6}} \over 2} \ \biggl( \
{\bigl( n_o + n_v + n_s + n_c \bigr)}^2 \ V_8 \ + \
{\bigl( n_o + n_v - n_s - n_c \bigr)}^2 \ O_8 \nonumber \\ &-& \
{\bigl( - n_o + n_v + n_s - n_c \bigr)}^2 \ S_8 \ - \
{\bigl( - n_o + n_v - n_s + n_c \bigr)}^2 \ C_8  \biggr) \quad , \nonumber
\\
{\tilde{M}}_{0b} \ &=& \ - 2 \ {{n_o + n_v + n_s + n_c } \over 2}  \ V_8  \quad
{}.
\label{open0b}
\ea
Now the open descendants are non-chiral in their closed sector but they are
chiral in their
open sector and thus, not surprisingly, there are tadpole conditions (see also
eq. (\ref{ktildetendim})) related to the
two ``unphysical'' scalars corresponding to $S_8$ and $C_8$.  The two resulting
conditions,
$n_o = n_v$ and $n_s = n_c$, dispose of all gauge anomalies.  The $V_8$
tadpole is again a physical one, corresponding to the dilaton.  Setting it to
zero would fix
to $64$ the total size of the gauge group.

In this class of models there are two more options for the Klein-bottle
projection, and
thus one may investigate the corresponding open sectors.  I shall discuss in
some detail
only the
models corresponding to $K^{\prime \prime}$ of eq. (\ref{kntendim}), since they
are more interesting. In general, when one moves from one projection to
another, complex
charges appear.  The need for this is easily seen by noticing that in this case
only $C_8$
can flow in the M\"obius amplitude.  This has an important consequence, namely
the gauge
vectors correspond to unitary groups, and thus carry pairs $n \bar{n}$ of
conjugate
charges.  Let me display the amplitudes, since this will make my remarks
clearer:
\ba
{A^{\prime \prime}}_{0b} \ &=& \  \bigl( n_1 {\bar{n}}_1 + n_2 {\bar{n}}_2
\bigr) \ V_8 \ +
\bigl( n_1 {\bar{n}}_2 + n_2 {\bar{n}}_1 \bigr) \ O_8 \nonumber \\ &-& \ \bigl(
n_1 n_2 +  {\bar{n}}_1
{\bar{n}}_2 \bigr) \ S_8
\ - \ {{{n_1}^2 + {n_2}^2 + {{\bar{n}}_1}^2 + {{\bar{n}}_2}^2} \over 2} \ C_8
\quad , \nonumber
\\
{M^{\prime \prime}}_{0b} \ &=& \ {{n_1 + {\bar{n}}_1 - n_2 - {\bar{n}}_2 }
\over 2}  \ C_8
\quad ,
\nonumber \\
{{\tilde{A}}^{\prime \prime}}_{0b} \ &=& \ {{2^{-6}} \over 2} \ \biggl( \
{\bigl( n_1 + {\bar{n}}_1 + n_2 + {\bar{n}}_2 \bigr)}^2 \ V_8 \ - \
{\bigl( n_1 - {\bar{n}}_1 + n_2 - {\bar{n}}_2 \bigr)}^2 \ O_8 \nonumber \\ &+&
\
{\bigl( n_1 - {\bar{n}}_1 - n_2 + {\bar{n}}_2 \bigr)}^2 \ S_8 \ - \
{\bigl( n_1 + {\bar{n}}_1 - n_2 - {\bar{n}}_2 \bigr)}^2 \ C_8  \biggr) \quad ,
\nonumber
\\
{{\tilde{M}}^{\prime \prime}}_{0b} \ &=& \ 2 \ {{n_1 + {\bar{n}}_1 - n_2 -
{\bar{n}}_2 }
\over 2 }
\ C_8
\quad .
\ea
It should be appreciated that the coefficients of $O_8$ and $S_8$ must both
vanish, since
their sign is necessarily incorrect! As anticipated,
however, this condition is directly implied by their relation to the dimensions
of the
fundamental representations of unitary groups.  Comparing with eq.
(\ref{ktildetendim}) shows that there is an additional tadpole condition for
the
``unphysical'' scalar mode in the
$C_8$ sector.  This yields a relation between the dimensions of the two unitary
groups,
namely
\be
 n_1 + {\bar{n}}_1 - n_2 - {\bar{n}}_2 \ = \ 64 \qquad ,
 \label{tadpolepp}
\ee
sufficient to ensure the cancellation of all gauge and gravitational anomalies,
as may
be seen making use of the results of ref. \cite{agw}. In this case one does not
have the
option to set to zero the dilaton tadpole, and the size of the gauge group is
therefore not
determined.  The option of relaxing the dilaton tadpole condition, considered
in ref.
\cite{bachas}, may lead to interesting progress in open-string string model
building.
The last class of open descendants may be constructed in a similar fashion, and
is left as
an exercise for the interested reader.  Details of the construction of open
descendants for large
classes of conformal models, including the factorization properties of disk
amplitudes
for minimal and $SU(2)$ $WZW$ models, may be found in refs. \cite{bps1}
\cite{wzw1} \cite{wzw2}.

Lower-dimensional models may be constructed in a similar fashion.  In
six-dimensions there
are large classes of chiral supersymmetric models with interesting gauge groups
that exhibit
a novel feature:
several antisymmetric tensor dispose together of gauge and gravitational
anomalies by a
generalized Green-Schwarz mechanism \cite{as}.  To date, however, chiral
four-dimensional
models are not understood to my own satisfaction, and apparently lead to
redundant
tadpole conditions that limit their gauge groups to a small size.

In conclusion, I have described in some detail the
construction of open descendants for some simple closed orientable models.
Some
general features of this procedure apply to arbitrary (left-right symmetric)
conformal models,
and endow them with one or more classes of open descendants.  Hopefully this
explicit derivation
will help interested
readers to get acquainted with some of the amusing properties of open-string
models that,
so far, we have been able to unveil.

\vskip 24pt
\begin{flushleft}
{\large \bf Acknowledgments}
\end{flushleft}

I am grateful to the Organizers for their kind invitation, and to
Massimo Bianchi, Gianfranco Pradisi and Yassen Stanev
for an enjoyable collaboration on many of the issues discussed in this
talk. This work was supported in part by C.N.R.S., France, and by
E.E.C. Grant CHRX-CT93-0340.

\vfill\eject

\end{document}